\documentclass{emulateapj}
\usepackage{graphicx}
\usepackage{natbib}

\singlespace

\shortauthors{Kogut et al.}
\shorttitle{Spectral Confusion for Cosmological C{\sc ii} Surveys}
\slugcomment{Accepted by The Astrophysical Journal}

\begin{document}

\title{Spectral Confusion for Cosmological Surveys 
of Redshifted C{\sc ii} Emission}

\author{ 
A. Kogut\altaffilmark{1},
E. Dwek\altaffilmark{1},
S. H. Moseley\altaffilmark{1}
}

\altaffiltext{1}{Code 665, Goddard Space Flight Center, Greenbelt, MD 20771}
\email{Alan.J.Kogut@nasa.gov}


\begin{abstract}
Far infrared cooling lines are ubiquitous features 
in the spectra of star forming galaxies. 
Surveys of redshifted fine-structure lines
provide a promising new tool
to study structure formation
and galactic evolution
at redshifts including 
the epoch of reionization
as well as 
the peak of star formation.
Unlike neutral hydrogen surveys,
where the 21 cm line is the only bright line,
surveys of red-shifted fine-structure lines
suffer from confusion generated by 
line broadening,
spectral overlap of different lines, 
and the crowding of sources with redshift.
We use simulations
to investigate the resulting spectral confusion
and derive observing parameters to minimize these effects
in pencil-beam surveys of red-shifted far-IR line emission.
We generate simulated spectra 
of the 17 brightest far-IR lines in galaxies,  
covering the 150 to 1300  $\mu$m wavelength region
corresponding to redshifts $0 < z < 7$,
and develop a simple iterative algorithm 
that successfully identifies the 158$\mu$m [C{\sc ii}] line and other lines.  
Although the [C{\sc ii}] line is a principal coolant
for the interstellar medium,
the assumption that the brightest observed lines
in a given line of sight
are always [C{\sc ii}] lines is a poor approximation
to the simulated spectra once other lines are included.
Blind line identification requires detection of fainter
companion lines from the same host galaxies,
driving survey sensitivity requirements.
The observations require moderate spectral resolution 
$700 < R < 4000$
with angular resolution  between 20\arcsec ~and 10\arcmin,
sufficiently narrow to minimize confusion 
yet sufficiently large 
to include a statistically meaningful number of sources.

\end{abstract}



\section{Introduction}
Line emission or absorption
is a promising probe of the high-redshift universe.
Considerable theoretical and instrumental effort
has been devoted
to the use of the redshifted 21 cm line 
of neutral hydrogen
in such a fashion
(see 
\citet{pritchard/loeb:2012}
for a recent review).
Comparable observations using 
atomic or molecular lines
at far-infrared wavelengths have seen less development.
Recent advances in far-IR instrumentation,
combining sensitive receivers
with large collecting area,
have led to the detection of far-IR lines
in individual sources at cosmological distances,
raising the prospects for
a future generation of far-IR line surveys
as a probe of the high-redshift universe.

Infrared fine structure lines are important coolants 
of the neutral and ionized gas phases 
of normal and starburst galaxies. 
They are easily excited, 
arise from the most abundant metals 
with low ionization potential, 
and are generally not affected by galactic or intergalactic attenuation.
The strongest line is the 
$^2P_{3/2} \rightarrow ~ ^2P_{1/2}$
fine-structure line of singly-ionized carbon 
at 157.74 $\mu$m rest wavelength. 
The ubiquity of carbon in the interstellar medium
combined with the relatively low ionization potential
and modest excitation temperature
make the [C{\sc ii}] line bright,
with as much as 0.1 -- 1\% of the total bolometric luminosity
of the host galaxy
emitted in this single line
\citep{crawford/etal:1985,
stacey/etal:1991,
wright/etal:1991,
lord/etal:1996}.
The intrinsic line brightness
and lack of significant attenuation
from the galactic or intergalactic medium
in the far infrared 
in turn
render the [C{\sc ii}] line visible over cosmological distances
\citep{maiolino/etal:2005,	
iono/etal:2006,			
maiolino/etal:2009,		
stacey/etal:2010,		
ivison/etal:2010,		
wagg/etal:2011,			
debreuck/etal:2011,		
venemans/etal:2012,		
wagg/etal:2012,			
neri/etal:2014,			
riechers/etal:2014,		
debreuck/etal:2014}.		

The ability to detect [C{\sc ii}] emission
to redshifts $z > 7$
makes the [C{\sc ii}] line
a promising candidate for cosmological surveys.
[C{\sc ii}] emission originates from multiple phases of the interstellar medium.
The combined emission from all phases
serves as a marker for source redshift
for surveys of large-scale structure
\citep{gong/etal:2011,
uzgil/etal:2014}.
The empirical relation between [C{\sc ii}] and the bolometric far-IR luminosity
allows the observed [C{\sc ii}] intensity
to serve as a tracer of star-formation activity
\citep{delooze/etal:2011,
delooze/etal:2014,
herrera-camus/etal:2014},
while the ratio of [C{\sc ii}] to other far-IR lines
probes physical conditions 
within different phases of interstellar medium of the host galaxies
\citep{abel:2006,
cormier/etal:2012,
croxall/etal:2012,
kapala/etal:2014}.

Several authors have discussed requirements 
for [C{\sc ii}] surveys at cosmological distances.
\citet{dacunha/etal:2013a}
estimate that a 500-hour observation
of the 2.4\arcmin ~diameter Hubble Deep Field
with the Atacama Large Millimeter Array
could yield 15 [C{\sc ii}] detections
over the redshift range $1 < z < 6$.
The sub-arcsecond angular resolution 
available for such interferometric observations
minimizes confusion from source superposition
within each synthesized beam,
but the large number of independent pointings
needed to map an area comparable to the HDF
minimizes the integration time spent on any individual source
so that only the brightest sources are detected.

Observations at coarser angular resolution
reduce the need for multiple pointings,
at the cost of 
introducing competing lines
from foreground sources.
Intensity mapping is one such technique,
using fluctuations 
in the spatial and spectral distribution
of line emission
from a superposition of sources 
to derive the underlying power spectrum
of the individual, unresolved sources
(see, {\it e.g.}, the discussion
in \citet{righi/etal:2008,
visbal/loeb:2010,
gong/etal:2011}).
Several such surveys have been proposed.
For example,
\citet{uzgil/etal:2014}
present a concept for [C{\sc ii}] intensity mapping
using a 25-beam grating spectrometer
each with beam width 10\arcsec,
while
\citet{silva/etal:2014}
propose a suite of single-beam spectrometers
each with 30\arcsec ~beam width.

The larger beam size anticipated for such surveys
facilitates mapping large areas
but requires additional analysis
to separate the target line emission
from the foreground created
by emission from other galaxies along the same line of sight.
Several authors discuss 
cross-correlation techniques 
to minimize foreground confusion
in line intensity mapping
\citep{righi/etal:2008,
visbal/loeb:2010}.
For a given target redshift,
the technique defines two observing wavelengths
corresponding to emission from two different rest-frame lines.
Maps of line emission
taken at each observing wavelength
contain spatially correlated structure
corresponding to the two lines emitted by each source galaxy.
The contribution from other lines 
at different rest-frame wavelengths
redshifted into the observing wavelengths
must originate from sources at different redshifts.
Spatial correlations from such distant sources will be weaker,
so that the cross-correlation
traces the power spectrum of the source galaxies
at the target redshift.

Intensity mapping surveys large volumes of the universe
to derive statistical estimates of structure formation,
but provides little information on individual sources.
In this paper,
we investigate the complementary technique
of direct line identification
in narrow-beam spectroscopic surveys.
If the [C{\sc ii}] line were the only observable far-IR line,
redshift identification would be straightforward.
In practice, however, 
the [C{\sc ii}] line is merely one of several cooling lines.
The problem of sorting through an observed spectrum
containing multiple lines
from multiple galaxies at different redshifts
presents both instrumental and analytical challenges.
We use a toy model 
of far-IR line emission
to evaluate a simple algorithm for line identification
and
estimate relevant instrumental parameters
(angular resolution, sensitivity, and spectroscopic resolution)
required to derive cosmologically interesting results.
The results are applicable
for spectral cleaning of intensity mapping surveys
as well as
pencil-beam surveys of more limited cosmological volume.

\begin{figure}[t]
\includegraphics[width=2.7in,angle=90]{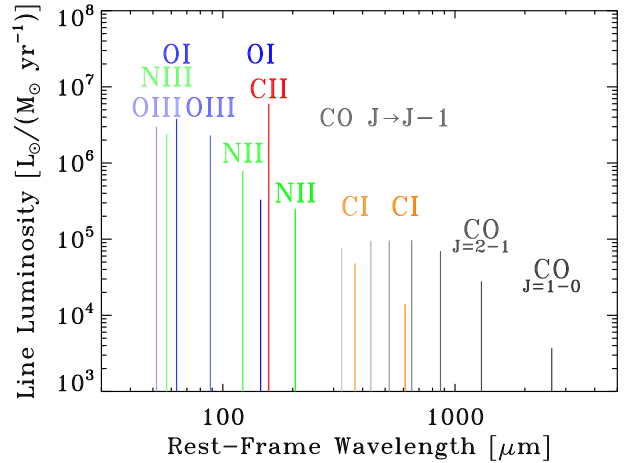}
\caption{
Rest-frame wavelength
and mean relative luminosity
for the far-IR lines used in this study
(see Table 1 in \cite{visbal/trac/loeb:2011}).
\label{line_fig} 
}
\end{figure}

\section{Methodology}
We use a simple model of the brightest far-IR lines
to simulate the observed line spectra from distant galaxies.
We assume that the total luminosity, $\ell(\nu_0)$ emitted in a single line
at rest frequency $\nu_0$
may be simply related to the
total infrared luminosity, $L_{IR}$ of a galaxy,
\begin{equation}
\ell(\nu) = f(\nu_0, L_{_{IR}})\, L_{IR} \, .
\label{f_def}
\end{equation}
The observed specific flux of a single line
from a set of IR galaxies 
at redshift $z$ is then given by
\begin{eqnarray}
F_{\nu}(\nu, z) &=& (1+z){ dV_c(z) \over 4 \pi d_L^2(z)}\ 	\\
&\times& \int_{L_1}^{L_2}\ dL_{_{IR}}\, \, \Phi(L_{_{IR}},z)\, 
f(\nu_0, L_{_{IR}})\, \phi_{\nu}(\nu')				\nonumber
\label{line_flux_def}
\end{eqnarray}
where 
$\Phi(L_{_{IR}},z)$ 
is the galaxy IR luminosity function,
$d_L(z)$ is the luminosity distance,
$\nu'=\nu(1+z)$
is the redshifted frequency, 
and
$\phi_{\nu}(\nu)$ 
is the intrinsic line profile, 
normalized so that  
$\int \phi_{\nu}(\nu)\, d\nu = 1$.
Ignoring the effects of gravitational lensing,
the total line intensity 
emitted from galaxies within a redshift interval
$z_0$  to the present ($z=0$) then becomes
\begin{eqnarray}
\label{obs_specific_intensity}
F_{\nu}(\nu) &=& {\Omega\over 4 \pi} \int_0^{z_1}  
\left|{cdt\over dz}\right|{dz\over 1+z} 		\\
 &\times&
\left[\int_{L_1}^{L_2}\ dL_{_{IR}}
\Phi(L_{_{IR}},z) f(\nu_0, L_{_{IR}})
\phi_{\nu}(\nu')\right] 				\nonumber
\end{eqnarray}
If the ratio of line intensity to IR luminosity is a constant,
Eq. (\ref{obs_specific_intensity}) simplifies to
\begin{equation}
\label{fnu3}
F_{\nu}(\nu) = {\Omega\over 4 \pi} f(\nu_0) L_{\star}  
\int_0^{z_1} n(z) \phi_{\nu}[\nu(1+z)]\left|{cdt\over dz}\right|{dz\over 1+z}
\end{equation}
where
\begin{equation}
\label{phi}
\int_{L_1}^{L_2} \Phi(L_{_{IR}},z)dL_{_{IR}} = n(z) L_{\star} \, ,
\end{equation}
$n(z)$ is the comoving number density of galaxies, 
and 
\begin{equation}
L_{\star} \equiv 
\frac{  \int\, L~\Phi(L)~dL }  {\int\, \Phi(L)~dL}
\label{Lstar_def}
\end{equation}
is their average IR luminosity.

We simulate the far-IR line spectra 
using a standard cosmology with
$h=0.7$,
$\Omega_m = 0.28$,
and $\Omega_\Lambda = 0.72$
\citep{hinshaw/etal:2013}.
We divide the comoving volume into redshift bins
and compute the total number of galaxies $N(z)$ within each bin
\begin{equation}
\label{dndz2}
{dN(z)\over dz} = n(z)\, {dV_c(z)\over dz}
\end{equation}
where $V_c(z)$ is the comoving volume at redshift $z$.
We use the resulting number density
to generate Monte Carlo realizations of the far-IR line spectra.
The background cosmology
fixes the mean number of galaxies 
within a fixed observing solid angle.
We assign each simulated galaxy
a redshift and IR luminosity,
using the \citet{saunders/etal:1990}
luminosity function
to obtain a representative luminosity distribution
within each redshift bin.

Each galaxy then contributes a set of far-IR lines.
Figure \ref{line_fig} shows the 
relative intensity of the rest-frame lines
used for this study.
We include the
[C{\sc ii}] line at 158 $\mu$m,
[C{\sc i}] at 610 and 371 $\mu$m,
[O{\sc iii}] at 88 and 52 $\mu$m,
[O{\sc i}] at 145 and 63 $\mu$m,
[N{\sc iii}] at 57 $\mu$m,
and 
[N{\sc ii}] at 205 and 122 $\mu$m,
as well as the 
CO ($J \rightarrow J-1$) series through $J=7$.
The total luminosity in each line
is related to that of the host galaxy
by the factor $f(\nu_0)$
(Eq. \ref{f_def}).
We set the mean value of $f$ for each line
using the ratios of values in 
Table 1 of \citet{visbal/trac/loeb:2011}.
For simplicity,
we do not explicitly calculate the star formation rate 
for each simulated galaxy,
but relate the comoving star formation rate $\dot \rho(z)$
to the IR luminosity density $L(z)$ as
\begin{equation}
{\cal L}(z) \approx {\cal A} \dot \rho(z)
\label{SFR_approx}
\end{equation}
where ${\cal L}(z)$ is the comoving IR luminosity density,  
$\dot \rho(z)$ is the comoving cosmic star formation rate (CSFR) 
in $M_\sun$~yr$^{-1}$~Mpc$^{-3}$, 
and ${\cal A} = 6.7 \times 10^9$ 
is a conversion factor for a Salpeter stellar initial mass function 
(Kennicutt 2012).

The relative luminosities 
from different lines within a source
are known to vary by roughly a factor of 10.
We account for such variation 
by
multiplying $f(\nu_0)$ for each line in each host galaxy
by a random number drawn uniformly from the range 
[0.3 -- 3].
We additionally account for the systematic variation
in line luminosity and line ratios
due to the 
increased CMB temperature
at higher redshift
\citep{dacunha/etal:2013b}.
We assume that the host galaxies are not spatially resolved
and assign a random rotational linewidth to each galaxy
uniformly drawn from the range [30 -- 300] km s$^{-1}$.
Each galaxy contributes multiple lines;
all lines from a single galaxy have the same rotational line broadening.
Finally, we accumulate the lines from all source galaxies
within the sampled volume
and redshift each line
to obtain a realization the observed far-IR spectrum.
The effects of instrument spectroscopic resolution
are discussed in $\S$3.2.
We ignore instrument noise
and continuum emission from the superposed host galaxies
to focus here on the problem of line identification.

\begin{figure}[b]
\includegraphics[width=3.5in,]{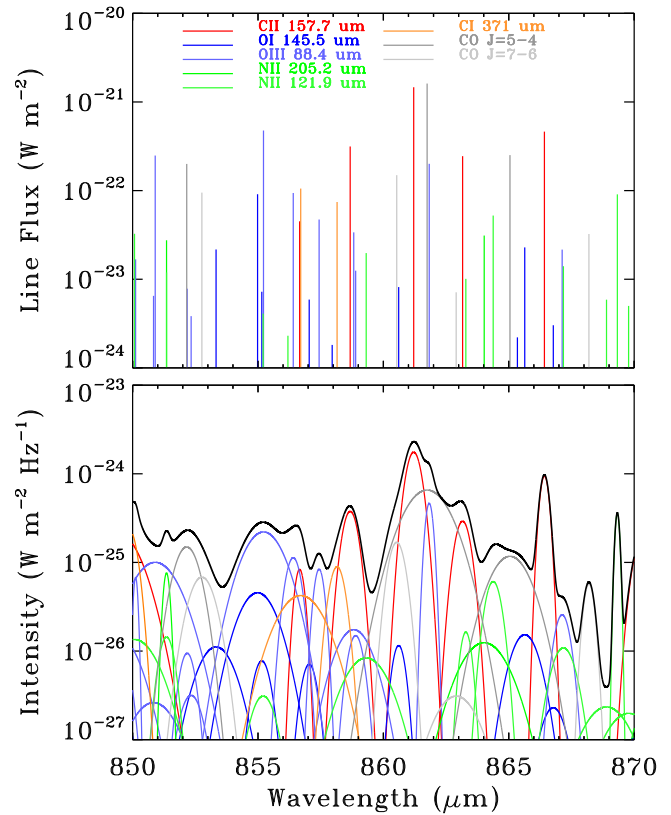}
\caption{Simulated spectrum with red-shifted lines.
The wavelength range corresponds to redshift 
$4.4 < z < 4.5$ for the [C{\sc ii}] line.
(Top) Individual line flux,
shown at the central wavelength
before rotational broadening.
Lines are color-coded by species
using the same color coding as Figure \ref{line_fig}.
48 lines from 48 individual galaxies at $0.6 < z < 8$ 
fall within 
the plotted wavelength range.
The assumption that the brightest lines
are always [C{\sc ii}] lines produces a high error rate,
with roughly a third of the brightest lines
originating from species other than [C{\sc ii}].
(Bottom) The same lines are shown after rotational broadening.
The superposition of rotationally broadened lines 
produces a complicated spectrum (black curve).
\label{sample_spectrum} 
}
\end{figure}

Figure \ref{sample_spectrum} shows a typical spectrum
with 2\arcmin ~angular resolution
at wavelengths corresponding to
redshifts $4.4<z<4.5$ for the [C{\sc ii}] line.
Two problems are evident.
The first problem is confusion.
Unlike radio observations,
where the redshifted 21 cm line is the only bright line,
a number of lines contribute to the observed far-IR spectra.
Only 16\% of the lines 
within the observed wavelength range
correspond to the targeted [C{\sc ii}] line,
with the remainder consisting of
``interloper'' lines
originating at both higher and lower redshifts
from species other than [C{\sc ii}].
The second problem is blending:
the superposition of rotationally broadened lines
produces a complicated spectrum.
The fainter lines merge to form a near continuum spectrum,
further hindering line identification.

\section{Line Identification}
Although the [C{\sc ii}] line is bright,
simply assuming that the brightest observed lines
within a single field
are always [C{\sc ii}] lines
produces a high error rate for line identification.
We quantify this by taking the simulated spectra
and selecting the brightest 10\% of lines
within a narrow range centered at different source redshifts.
We may then count how many of the brightest lines
in each range
are in fact [C{\sc ii}] lines
as opposed to interloper lines from sources at other redshifts.
The probability that 
a given line
within the brightest 10\%
is in fact a [C{\sc ii}] line
falls from 0.8 for observations near 315 $\mu$m ($z_s = 1$)
to 0.4 for observations at 1100 $\mu$m ($z_s = 6$).

We address the problem of blind line identification
using a version of the radio-astronomical {\tt CLEAN} algorithm.
Although the brightest lines are not necessarily [C{\sc ii}] lines,
the [C{\sc ii}] lines are generally bright.
This suggests an iterative process.
We first select the brightest line within the observed spectrum.
Assuming that this ``parent'' line is [C{\sc ii}]
fixes the corresponding source redshift $z_s$.
We then predict the wavelengths
at which other far-IR lines originating from the same source
should be observed,
\begin{equation}
\lambda_{\rm pred} = \lambda_{\rm rest} ~ (1 + z_s) ,
\label{CLEAN_eq}
\end{equation}
and examine the spectra at those wavelengths.
If the expected ``companion'' lines are found,
the redshift identification is confirmed
and
both the candidate [C{\sc ii}] line
and the detected companion lines
are flagged as known
and removed from the list of lines to search.
If no companion lines are found,
we repeat the test 
assuming that the parent line
is the next-most-common far-IR line
(typically a CO line)
and continue down the list of far-IR lines
until
we identify a parent/companion combination
or
exhaust the list of potential parent lines.
If no companion lines are found,
the bright parent line is flagged as unknown
and removed from the list of possible parent lines,
although it may later be identified as a companion line.
We then proceed to the brightest remaining un-flagged line
and iterate until all lines are tested.

\begin{figure}[t]
\includegraphics[width=3.5in]{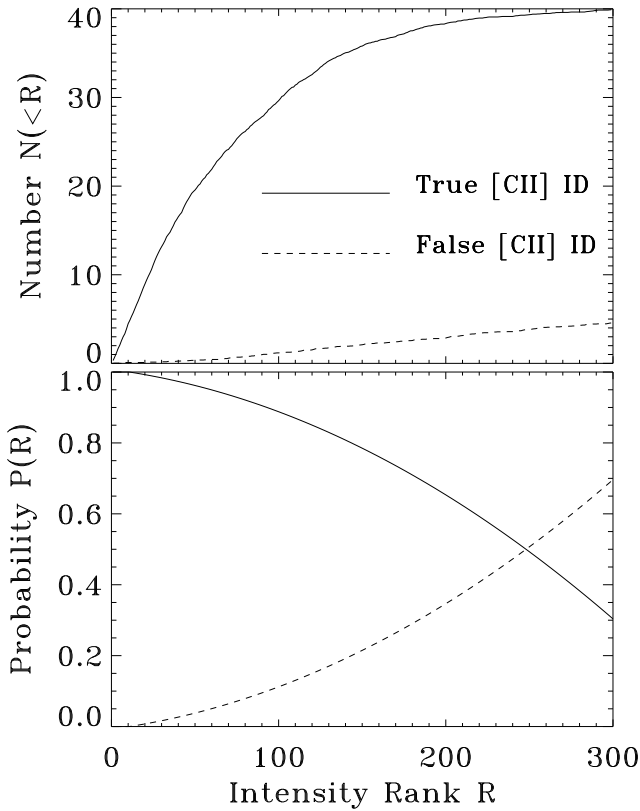}
\caption{Results of the {\tt CLEAN} algorithm
compared to simulation inputs
for a simulated spectrum 
from 580 $\mu$m to 1160 $\mu$m
containing 390 peaks
(see text).
(Top) Cumulative number of
correct (solid line) and incorrect (dashed) 
[C{\sc ii}] line identifications
as a function of the relative line intensity,
sorted from brightest ($R=0$)
to faintest ($R=300$).
(Bottom) Probability
that the $R^{\rm th}$-brightest peak
in the observed spectra
will be correctly identified.
The number of false identifications
is small for the brightest half of the observed lines.
\label{error_fig} 
}
\end{figure}

\subsection{Efficiency and Error Rate}
We test the algorithm using Monte Carlo realizations 
of the observed spectra
and compare the derived identifications
to the known Monte Carlo inputs
to determine the detection efficiency and error rate.
Figure \ref{error_fig} shows the results
obtained from a single 2\arcmin ~beam
observing one octave in wavelength
from 580 $\mu$m to 1160 $\mu$m,
corresponding to [C{\sc ii}] redshifts
$2.7 < z_s < 6.3$.
The input spectrum contains
1600 spectral lines from 550 distinct galaxies,
including 260 [C{\sc ii}] lines.
After rotational broadening,
the observed superposed spectrum
contains 390 peaks\footnote{
Working with noiseless simulations,
we define a ``peak'' as any spectral bin
brighter than two neighbors on either side.
The effects of coarser instrumental resolution
are discussed in $\S$ 3.2},
110 of which result from a [C{\sc ii}] line.
Blending is significant:
of the 390 peaks in the superposed spectra,
290 are dominated by a single line
and
100 are blends of multiple lines.
The un-blended peaks 
comprise 175 unique host galaxies.

The line identification algorithm starts with the brightest such peak
and works to progressively fainter peaks.
The algorithm correctly identifies 200 lines
corresponding to 80 unique parent redshifts
and 120 companion lines.
40 of the correctly identified parent lines 
result from [C{\sc ii}] emission,
with the remainder primarily from
lower-redshift CO lines.
The detection efficiency is thus
15\% of the 260 [C{\sc ii}] lines input to the simulation,
or
36\% of the 110 potentially detectable [C{\sc ii}] lines
in the superposed spectra.

The algorithm correctly identifies half
of the 390 peaks in the superposed spectra.
The remaining 190 peaks
are either incorrectly identified (160 peaks)
or not identified (30 peaks).
False identifications 
typically result from blended lines
and
are spread uniformly across 
all 17 candidate lines
(i.e. with 5--10 false identifications
for each line in Figure \ref{line_fig}).
The error rate is a function of relative line intensity,
with the probability of a correct identification
near unity for the brightest 10--15 lines
and remaining above 75\%
for lines in the brightest half of the observed spectrum
(30--40 lines for the parameters in Figure \ref{error_fig}).
Lines in the faintest third of the observable spectrum
are more likely to generate a false identification
than a true identification.
Restricting the algorithm to searching only the brightest half
of the observed set of lines
reduces the number of correctly-identified [C{\sc ii}] lines 
from 40 to 38
while yielding only 3 false identifications,
for an error rate of 7\%.

Incorrect line identification
results from the chance spectral alignment
of one line from one galaxy
with a different line from a different galaxy.
The spectral density of lines
increases at fainter intensities
(Fig \ref{sample_spectrum}),
increasing the probability for such alignment.
Restricting potential ``parent'' lines
to the brightest half of observed lines
largely eliminates this problem
without sacrificing correct identifications.
Incorrectly identified lines
may also be flagged
through anomalous line ratios,
although in this work we do not apply such a line ratio test.

\begin{figure}[b]
\includegraphics[width=2.7in,angle=90]{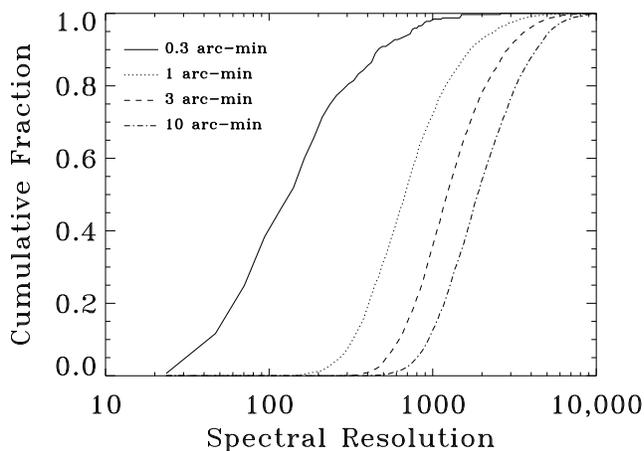}
\caption{Fraction of adjacent peaks
in the simulated spectra
that can be resolved at spectral resolution $R$.
Results are shown
for angular resolution
0.3, 1, 3, and 10 arc-minutes (left to right).
Observations at modest spectral resolution
($R \sim 3000$)
resolve as much of the blended spectrum as possible.
The superposition of rotationally broadened lines
from multiple sources
creates an effective ceiling at $R \sim 5000$.
Observations at higher spectral resolution
do not distinguish additional lines.
\label{spectral_res} 
}
\end{figure}

\subsection{Spectral Resolution}
The {\tt CLEAN} algorithm
identifies [C{\sc ii}] lines
by searching for predicted ``companion'' lines
among the set of observed spectral lines.
As the number of lines increases,
confusion from overlapping or blended lines
can create errors in the position of the fitted line centers
to limit the efficiency of the algorithm.
We use simulated spectra to assess the
instrumental parameters
required to mitigate line confusion.

The number of lines
depends on the number of galaxies 
within a spatially unresolved observation.
Decreasing the angular resolution
increases both the number of galaxies observed
and the
spectral density of emission lines.
Observing more spectral lines
in turn
increases the probability
that lines 
from different source redshifts
will overlap in wavelength,
requiring higher spectral resolution
to distinguish individual lines.

We quantify the spectroscopic resolution
as a function of the angular resolution.
For a given angular resolution,
we simulate the galaxy population
and generate the superposed spectra
(including rotational broadening)
from all sources
using the species in Figure \ref{line_fig}.
We identify all peaks in the resulting (noiseless) spectra
and compare the separation $\Delta \lambda$
between adjacent peaks
to the wavelength $\lambda$
to determine the spectral resolution 
$R = \lambda / \Delta \lambda$
needed to distinguish the two as separate lines.

Figure \ref{spectral_res}
shows the fraction of the observed spectral peaks
whose separation $\Delta \lambda$ 
from the neighboring peak
is greater than than $2 \lambda / R$,
where the factor of 2 accounts for 
the need to observe two distinct peaks.
At low spectral resolution,
the lines are heavily blended
so that only the few brightest lines
can be distinguished.
As the instrument spectral resolution increases,
progressively more of the lines can be distinguished.
The spectral resolution needed to distinguish separate peaks
depends on the angular resolution:
decreasing the angular resolution
reduces the number of galaxies observed,
producing correspondingly greater separation
between adjacent peaks in the observed spectra.
Spectral resolution $R = 700$
(velocity resolution 430 km s$^{-1}$)
successfully resolves 90\% of the peaks in the blended spectra
for a 20\arcsec ~resolution.
A higher resolution
$R = 4000$ 
(75 km s$^{-1}$)
is required
to resolve  90\% of the peaks
for 10\arcmin ~resolution.
Rotational line widths of a few hundred km~s$^{-1}$
effectively limit the resolution to values $R < 5000$
independent of the angular resolution.
Increasing the spectral resolution above $R = 5000$
does not distinguish additional lines,
as the observed spectra
are then limited by intrinsic source blending
and not by the instrument.

\begin{figure}[t]
\includegraphics[width=2.5in,angle=90]{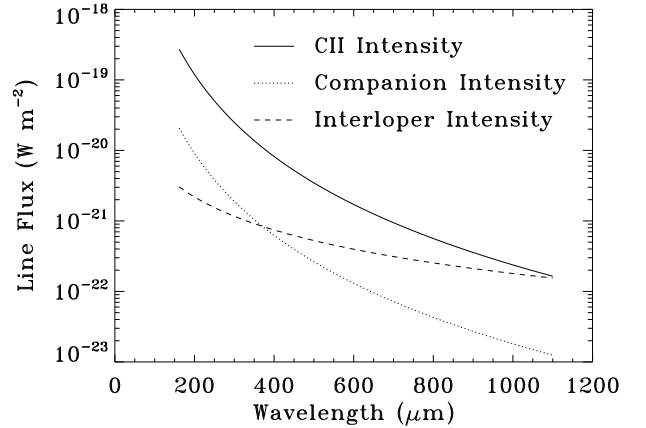}
\caption{Flux from the red-shifted [C{\sc ii}] line
(solid line)
as a function of observing wavelength.
The mean flux of the brighter companion lines
from each [C{\sc ii}] source is shown as a dotted line,
while the dashed line
shows the flux of
competing emission from unrelated interloper lines.
\label{sensitivity_fig} 
}
\end{figure}

\subsection{Sensitivity}
The line identification algorithm requires sufficient 
instrumental sensitivity
to detect both the [C{\sc ii}] line
as well as one or more companion lines.
Figure \ref{sensitivity_fig}
shows the typical line flux
as a function of observing wavelength.
For each [C{\sc ii}] source in the simulation,
we compute the [C{\sc ii}] line intensity
as well as the mean intensity
of the four companion lines
([O{\sc i}] 146 $\mu$m,
[O{\sc iii}] 88$\mu$m,
[N{\sc ii}] 205$\mu$m,
and
[N{\sc ii}] 122$\mu$m)
most often used by the algorithm
for successful [C{\sc ii}] identifications.
To the extent that physical conditions 
in the interstellar medium remain constant,
the ratio of companion to [C{\sc ii}] line intensity
remains constant
so that the observed line fluxes simply fall 
at the longer observing wavelengths
in accordance with the higher source redshift.
Interloper lines, however,
include sources at multiple redshifts
unrelated to the [C{\sc ii}] source redshift
and thus fall more slowly with wavelength.

The companion lines
are roughly a factor of 10 fainter than the [C{\sc ii}] line.
Observations using the {\tt CLEAN} algorithm to determine source redshift
must retain sufficient sensitivity
to observe these fainter companion lines
in addition to the targeted [C{\sc ii}] lines.
Interloper lines become brighter than the companion lines
for wavelengths $\lambda > 350~\mu$m
corresponding to [C{\sc ii}] source redshift $z_s > 1$.
Observations at longer wavelengths
targeting higher source redshifts
require sufficient spectral resolution
to distinguish the companion lines
from the brighter interloper lines
(Figure \ref{spectral_res}).

The requirement to distinguish 
at least one identifiable companion line 
for each [C{\sc ii}] line
implies that the sensitivity and integration times
for a blind survey
are determined by the fainter companion lines
and not the targeted [C{\sc ii}] lines.
To reduce the total observing time
needed to map a given field,
surveys are likely to employ
multiple spectrometers
simultaneously observing independent lines of sight.
Such an arrangement has been envisioned 
for future surveys
\cite{silva/etal:2014,
uzgil/etal:2014}

\section{Pencil-Beam Survey}
The ubiquity and intensity of line emission from ionized carbon
makes the [C{\sc ii}] line
a promising probe of the high-redshift universe.
We may thus anticipate
future surveys using the [C{\sc ii}] line
for redshifts $z < 10$
to probe
large scale structure,
galaxy evolution,
and the epoch of reionization
\citep{suginohara/etal:1999,
righi/etal:2008,
visbal/trac/loeb:2011,
gong/etal:2012,
silva/etal:2014,
uzgil/etal:2014}.
Instrumental requirements
for such a survey, however, are not clear.
Observations from ground-based platforms
(e.g. ALMA or PdBI)
allow large collecting areas
but are limited to the available atmospheric windows.
Even within these windows,
atmospheric emission
is 6 orders of magnitude 
brighter than the astrophysical foregrounds.
The resulting photon noise
limits sensitivity,
requiring integration times of 500 hours
to detect a handful of sources
within a cosmologically interesting volume
\citep{dacunha/etal:2013a}.

Observations from a balloon or space platform
can mitigate atmospheric emission
at the cost of limiting the size of the
beam-forming optics.
The smaller collecting area
reduces sensitivity to individual unresolved sources,
while the coarser angular resolution
increases confusion from multiple sources
within each angular resolution element.
\citet{uzgil/etal:2014}
discuss intensity mapping from a balloon platform
as one way to cope with these problems.
Here we consider the complementary approach
of a pencil-beam survey
from a small millimeter-wave telescope.
A pencil-beam survey
allows a single instrument
to measure multiple galaxies
within each line of sight,
using spectroscopy to separate individual sources
and determine source redshifts.

\begin{figure}[t]
\includegraphics[width=3.5in]{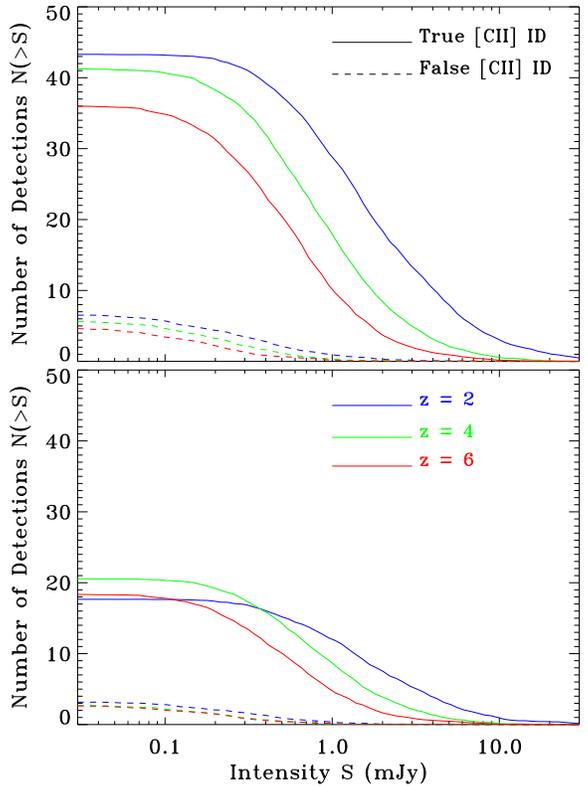}
\caption{Number of identified [C{\sc ii}] lines
as a function of line intensity.
Solid lines show the 
cumulative [C{\sc ii}] number counts derived from simulations
of a single 2\arcmin ~beam
observed over one octave in wavelength
centered at redshifts 2 (top curve, blue), 
4 (green), or 6 (bottom curve, red).
Dashed lines show the number of false identifications.
Companion lines are not shown.
(Top) Total number of [C{\sc ii}] detections 
over the full wavelength band.
(Bottom) Subset of detected lines 
with redshift lying within 20\% of the target redshift.
\label{mjy_fig} 
}
\end{figure}

Figure \ref{mjy_fig}
shows the number of [C{\sc ii}] lines
that could be identified
within a single pointing
for a noiseless instrument.
We simulate a single 2\arcmin ~beam
observed with spectral resolution $R = 3000$
over one octave in wavelength
and use the {\tt CLEAN} algorithm 
to identify lines within the resulting spectra.
We examine 3 cases
for spectral bands
300--600 $\mu$m,
500-1000 $\mu$m,
and
700--1400 $\mu$m
corresponding to targeted [C{\sc ii}] redshifts
$z$ = 2, 4, and 6.
The upper panel shows the cumulative number count
$N(>S)$
for all [C{\sc ii}] lines 
brighter than threshold $S$
that were identified by the algorithm.
The broad spectral band
allows the number counts 
to include sources over a range of redshifts.
The bottom panel
shows the number counts
for the subset of detected lines
within 20\% of the target redshift.
A single pointing 
at 5$\sigma$ threshold of 0.3 mJy
can potentially identify
20--40 redshifted [C{\sc ii}] lines
(plus a comparable number of companion lines)
within a 2\arcmin ~beam.

\begin{figure}[t]
\centerline{
\includegraphics[width=2.7in,angle=90]{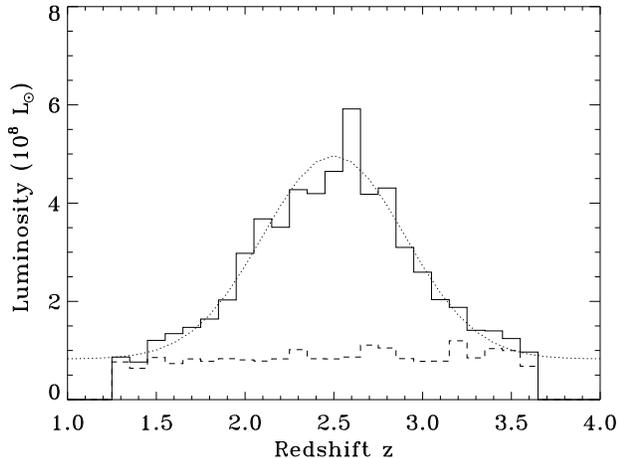} }
\caption{Galaxy luminosity distribution
derived from a simulated pencil-beam survey
observing the [C{\sc ~ii}] line
over one octave in wavelength
at spectral resolution $R = 3000$.
The recovered luminosity distribution (solid line)
accurately follows the toy model (dotted curve).
The dashed line shows the recovered luminosity distribution
for a model without luminosity evolution.
\label{luminosity_evolution} 
}
\end{figure}

We illustrate the potential utility of a pencil-beam survey
using a toy model of the star formation history.
We compare a model
with no luminosity evolution
to a second model
for which the typical luminosity
increases by a factor of 5 
at redshift $z_p$.
Figure \ref{luminosity_evolution} shows the result.
The recovered luminosity distributions
accurately reflect the toy model luminosity evolution.
Such a pencil-beam survey
easily distinguishes a model
in which the luminosity peaks at $z_p \sim 2.5$
from a model with no evolution.

\section{Conclusions}
The [C{\sc ii}] line is a principal coolant of the interstellar medium
and is observable at cosmological distances.
However, the [C{\sc ii}] line
is observed against
a forest of competing lines
from sources at multiple redshifts.
We use Monte Carlo simulations of the galaxy distribution
at redshifts $0 < z < 9$
to generate realizations of the far-IR spectra
including emission
from 17 of the brightest far-IR lines.
The resulting spectra provide a test bed
to evaluate the use of the [C{\sc ii}] line
as a cosmological probe in blind pencil-beam surveys.

The superposition of 
rotationally-broadened lines
from
multiple galaxies at different redshifts
creates a complex spectrum.
The simplest assumption,
that the brightest lines
are always [C{\sc ii}] lines,
is at best a coarse approximation to the simulated spectra.
The fraction of [C{\sc ii}] lines
among the brightest 10\% of all observed lines
falls from 0.8 for observations near 315 $\mu$m ($z_s = 1$)
to 0.4 at 1100 $\mu$m ($z_s = 6$).
Line identification based solely on relative intensity
produces a high error rate.

A variant of the radio-astronomical {\tt CLEAN} algorithm 
yields superior results.
The algorithm selects the brightest line
from an observed set of spectral lines,
tentatively identifies the line as [C{\sc ii}],
and uses the resulting source redshift
to predict the observed wavelengths
for redshifted line emission
from other bright far-IR lines from the same source.
If the predicted lines are observed,
the source identification is confirmed
and the [C{\sc ii}] line and redshifted companion lines
are flagged and removed from the list of unknown lines.
If the predicted lines are not observed,
the source identification is not confirmed
and the bright source line
remains on the list of unknown lines
for possible later identification 
as a companion line for a source at a different redshift.
The algorithm then proceeds to the next brightest remaining line
and iterates until all lines have been processed.

A typical spectrum for 2\arcmin ~angular resolution
contains some 1600 spectral lines within one octave,
originating from 550 different galaxies.
The superposition of rotationally broadened lines
reduces this to roughly 390 distinct peaks
in the observed spectrum,
110 of which result from a single un-blended [C{\sc ii}] line.
The line identification algorithm correctly identifies 
40 of the 110 [C{\sc ii}] line peaks
for an efficiency of order 36\%
(not counting companion lines identified from the same host galaxies).
Chance alignments of lines
lead the algorithm to incorrectly identify 3 peaks
as [C{\sc ii}] lines;
however,
the incorrect identifications
result almost exclusively
from fainter candidate peaks.
Restricting the algorithm to the brightest third 
of the observed peaks
(corresponding roughly to threshold 0.3 mJy)
increases the percentage of correct identifications above 90\%.
Correct line identification 
requires detection of both the targeted ``parent'' line
and one or more ``companion'' lines from the same source.
Survey sensitivity thresholds 
are thus determined by the expected intensities
of the companion lines
and not solely by the brighter [C{\sc ii}] line.

The ability to distinguish individual peaks in the superposed spectra
depends on the spectral resolution and angular resolution.
Decreasing the angular resolution
increases the number of lines in the observed spectra,
requiring higher spectral resolution
to minimize blending of neighboring lines.
The required resolutions are modest.
Spectral resolution $R = 700$
resolves 90\% of the peaks
for 20\arcsec ~angular resolution,
while
$R = 4000$
resolves 90\% of the peaks
for 10\arcmin ~angular resolution or coarser.

The [C{\sc ii}] line 
is a promising cosmological probe.
The large number of potential sources
observed within a single pencil beam
combined with an
efficient and accurate line identification algorithm
allow rapid characterization
of the galaxy luminosity distribution
to redshifts encompassing reionization.
A simulated pencil-beam survey
demonstrates
the ability to distinguish 
toy models of galaxy evolution
within a single angular resolution element.
The required angular resolution,
sensitivity,
wavelength range,
and spectroscopic resolution
appear achievable,
suggesting that far-IR lines
may soon provide a new tool
to map the large-scale galaxy distribution and evolution
for redshifts $0 < z_s < 10.$

\acknowledgements
We thank D. Leisawitz for encouraging development
of the simulations.  
Support for this research came from NASA's
Science Innovation Fund.


\end{document}